\documentstyle[preprint,aps,tighten]{revtex}

\def\be{\begin{equation}}
\def\ee{\end{equation}}
\def\br{\begin{eqnarray}}
\def\er{\end{eqnarray}}

\newcommand{\half}{\frac{1}{2}}
\def\q{$q$-}
\def\dag{{\dagger}}

\begin{document}
\draft
\title{Transformations of  q-boson and q-fermion  algebras }
\author{ P. Narayana Swamy $^a$}
\address{
Physics  Professor Emeritus, Southern Illinois University,
Edwardsville, IL 62026, USA}
\maketitle

\begin{abstract}

We investigate the algebras satisfied by q-deformed boson and fermion oscillators, in particular the transformations of the algebra from one form to another. Based on a  specific algebra proposed in recent literature, we show that the algebra of deformed fermions can be transformed to that of undeformed standard fermions. Furthermore we also show that the algebra of q-deformed fermions can be transformed to that of undeformed standard bosons. 

\end{abstract}

\vspace{0.5in}

\noindent  $^a$ electronic address: pswamy@siue.edu

\vspace{0.5in}


The study of quantum groups have led to theories of \q deformed boson and fermion oscillators based on a deformation \cite{Bieden} of the standard algebra of boson and fermion creation and annihilation operators.  It has recently been shown that the thermostatistics of \q deformed bosons can be realized in a self-consistent manner if the statistical mechanics of such systems is developed by using the  \q calculus  based on Jackson derivatives \cite{Alpns}. Recent interest has focussed on the proposition \cite{Partha,Jing} that the \q deformation of fermion oscillators may or may not reduce to the algebra of ordinary fermions obeying Pauli principle depending on the particular algebra that is postulated. Chaichian {\it et al}  \cite{Chai} have pointed out the importance of transformations which transform such algebras from one form to another, while  investigating the connection between   \q oscillator statistics and fractional statistics. Specifically Chaichian {\it et al } argue that there is a special formulation of the \q deformed fermion algebra which cannot be reduced to the algebra  of ordinary fermions in this manner.  In view of the importance of such transformations which map one kind of algebra  to another, it is quite important  to investigate  the nature of such transformations in the most general context. 

In this note we study the general properties of such transformations generated by operator functions of the number operator and conclude that transformations from deformed fermions  to ordinary fermions or  ordinary bosons is possible.

Let us begin with the standard  boson algebra defined by the commutation relations
\be
a_0 a^\dag_0 - a^\dag_0 a_0 = 1 , \quad  [N,a_0]= -a_0, \quad  [N, a^\dag_0 ]= a^\dag_0, \quad [\, a_0, a_0\, ] =  [\, a^{\dag}_0, a^{\dag}_0\, ] =0\, , 
\label{1}
\ee
where the number operator is $N=a^\dag_0a_0$. We have suppressed the particle index for simplicity. The  \q deformed bosons \cite{Bieden} can be introduced by the algebra 
\be
a a^\dag - qa^\dag a = q^{-N}, \quad  [N,a]= -a, \quad  [N, a^\dag]= a^\dag,
\label{2}
\ee
where $q$ is a positive real number, $N$ is the \q boson number operator and  $a, \, a^\dag, \, $ are the annihilation, creation operators respectively. The number operator is given by 
\be
N = \frac{1}{2 \ln q}\ln \left ( \frac{a a^{\dag} + q^{-1}a^{\dag}a}   { a a^{\dag} + q a^{\dag}a }  \right )
\ee
and depends on $q$.
We note that $N= a^\dag a$ only in the limit $q \rightarrow 1$ when the algebra in Eq.(\ref{2}) reduces to that in Eq.(\ref{1}).  The Fock representation of this algebra for the eigenstates $|n \rangle $  of the number operator is enabled by the construction
\be
|n \rangle = \frac{(a^\dag)^n}{\sqrt{  [n] ! }   }|0 \rangle, \quad a|0 \rangle =0, \quad N|n \rangle = n |n \rangle\, .
\label{3}
\ee
Here, $[n]! \equiv [n] \, [n-1]\, \cdots [1], \; [0]!=1 $ and [n] is the basic number defined by
\be
[x]= \frac{q^x - q^{-x}} {q-q^{-1}}.
\label{4}
\ee
In this Fock space,  the following relations are easily obtained: $a^\dag a = [N],  \;  a a^\dag = [N+1]\, $. 

It is of importance to discuss possible transformations on the creation and annihilation operators which can transform the algebra from one form to another.  For instance \cite{Chai},  
the transformation  defined by
\be
a_1 = q^{N/2}a, \quad  a_1^\dag = a^\dag q^{N/2}\, ,
\label{5}
\ee
introduces   new operators which  satisfy the algebra 
\be
a_1 a_1^\dag - q^2 a_1^\dag a_1 =1.
\label{6}
\ee
In this representation,  we find $a_1^{\dag}a_1 = q^{N-1} [N]$ and $a_1 a_1^{\dag}= q^N [N+1]$.  

We shall now consider a general transformation of the \q boson algebra by introducing
\be
A= f^{\half}a \, , \quad a^{\dag} = a^{\dag} f^{\half}\, ,
\ee
where $f = f(N)$ can be taken to be a general function of $N$.
In order to study such  transformations in a general context, let us first develop a useful construction, a lemma. Consider  an  operator $f(N)$  which is a function of the number operator. We can establish the following well-known fundamental relation \cite{Partha,Jing} ,  which we shall refer to as the commutation property:
\be
f(N) a = a f(N-1), \; f(N) a^\dag = a^\dag f(N+1),
\label{7}
\ee
which follows from Eq.(\ref{2})for any polynomial $f(N)$.  Eq.(\ref{6})  is then easily established using this construction. Later we shall  prove this commutation property for a general class of functions.

Let us now  consider a different transformation  $a_2= f^{\half}a, \; a_2^\dag = a^\dag f^{\half}$, where $f$ stands for a real polynomial function $f(N)$, satisfying the commutation property. We may now ask under what conditions  the transformed operators satisfy the algebra $a_2 a_2^\dag - a_2^\dag a_2 =1$. We see by using the commutation property that this  requires the necessary condition
\be
[N+1] f(N) - [N] f(N-1)=1.
\label{8}
\ee
One solution is $f(N) =  (N+1) / [N+1] $. Consequently,  the transformation
\be
a= \left ( \frac{[N+1]}{N+1}\right )^{\half}a_2, \quad a^\dag = a_2^\dag \left ( \frac{[N+1]}{N+1} \right )^{\half}
\label{9}
\ee
transforms it to the undeformed standard boson algebra 
\be
a_2 a_2^\dag - a_2^\dag a_2 =1, \quad [N,a_2]= - a_2, \quad  [N, a_2^\dag ]= a_2^\dag \, .
\ee
This is identical with the  algebra of  $a_0, a^\dag_0$, Eq.(\ref{1}).  That the \q deformation of bosons defined by Eq.(\ref{2}) can be transformed away in this manner  is   a known result \cite{Chai}. 

Since such  transformations are quite interesting and  useful in the analysis of \q deformed boson and fermion oscillators, we must investigate the general validity of  Eq.(\ref{7}) in more detail.  In what follows, $a$ referring to Eq.(\ref{7}), represents a generic annihilation operator obeying   $ [N,a]= -a, \quad  [N, a^\dag]= a^\dag $, and could represent either 
the \q boson algebra or the \q deformed fermion algebra. First,  the commutation property is clearly valid for $N$, as is evident from Eq.(\ref{2}). The proof then extends to $ N^2$ or  for  any monomial $N^r$ and hence for a polynomial function of $N$.   It is easy to verify the property  for  the case when $f =1/N$ and then for any negative power of $N$. We can then extend the proof to the case when $f(N)=q^N$ by observing that 
\br
q^N a &=& e^{N \ln q}a = \left \{ 1 + N \ln q + \frac{N^2}{2!}(\ln q)^2 + \cdots     \right \}a \\
&=& a \left \{ 1 + (N-1) \ln q + \frac{(N-1)^2}{2!}(\ln q)^2 + \cdots     \right \} = a q^{N-1}.
\er
Similarly, the property is valid for $q^{-N}$. In all these cases, it can be easily  proved that the transformed operators $a, a^\dag$ satisfy the commutation relations $[N,a] = -a$ and $[N,a^\dag ]= a^\dag$. For instance, in the case of the transformation used in Eq.(\ref{9}), due to Eq.(\ref{2}), we find 
\be
[N,a_2]= [N, f^{\half} \, a]=\,  f^{\half} [N,a]=\, -a_2 \, .
\ee
Next we turn our attention to more general functions. The commutation property is true for the basic number $[N]$ since it contains monomials. Alternatively we can readily verify it for this case since  $[N+1] a= a a^\dag a= a [N]$. This can be readily extended to power functions such as $[N]^r$ and to fractional powers of the basic number such as $[N]^{\half}$ and  $[N]^{-\half}$ and hence  functions of the form $[N+1]/(N+1)$. We can next extend the proof to the operator function  $(-1)^N q^N$ by observing 
\br
(-1)^N q^N f  &=& e^{i \pi N} e^{N \ln q}f = \left ( 1 + N(\ln q + i \pi) + \frac{N^2}{2!} (\ln q + i \pi )^2 + \cdots    \right )f \\
& =& f(-1)^{N-1}q^{N-1}\, .
\er
This enables us to extend the validity of  the commutation property  for $[N]^F$, the \q deformed fermionic basic number introduced in ref. \cite{Chai} and hence to operators such as  $[N+1]^F/(N+1)$.
We have thus proved the validity of the commutation property, Eq.(\ref{7}), for a large class of operator functions of the number operator $N$. We shall now proceed to derive some interesting consequences  of such transformations.

We begin with the algebra of the standard fermions defined by
\be
b_0 b^\dag_0 + b^\dag_0 b_0 = 1, \quad [N,b_0] = -b, \quad [N, b^\dag_0] = b^{\dag}_0,  
\quad \{ \, b_0, b_0 \, \} = \{\, b_0^{\dag}, b_0^{\dag} \, \} =0  \, , 
\label{16}
\ee
where  the fermion number is given by $N= b^\dag_0 b_0$. It follows from the above relations that $b_0^2 = 0, \, (b^\dag_0)^2 =0 $ and thus Pauli exclusion principle is contained in the algebra.  Next let us proceed to investigate  the \q deformation theory based on the \q fermion algebra \cite{Chai}  by introducing  
\be
bb^\dag + q b^\dag b=q^{-N}, \quad    [N, b] = -b\, , \quad [N, b^{\dag}] = b^{\dag}\, ,
\label{17}
\ee
where $N$ is the number operator of the \q fermions. The \q fermionic basic number is defined by
\be
[x]^F = \frac{q^{-x} - (-1)^x q^x}{q+q^{-1}}\, ,
\ee 
which is quite different from the basic number, Eq.(\ref{4}),  introduced in the theory of \q bosons.  The eigenstates of the number operator in this theory are 
\be
|n \rangle = \frac{(b^\dag )^n}{\sqrt{[n]^F !   }}.
\ee
This algebra plays a major role in the theory  \cite{Partha,Chai} of generalized \q deformed fermions, whose characteristic feature is that states with more than one fermion are allowed, thus going beyond the Pauli exclusion principle:   $b^2=0, \; (b^\dag )^2=0$ is realized only in the limit $q \rightarrow 1$ and not true generally when $q \not= 1$. We then have
\be
b^\dag b = [N]^F\, , \quad b b^\dag = [N+1]^F\, .
\label{20}
\ee
We now ask if we can transform this algebra to the algebra of ordinary undeformed fermions.  For this purpose,  consider the transformations
\be
b \rightarrow  \; b_1= T(N) b, \quad b^\dag \rightarrow  \; b_1^\dag = b^\dag T(N)\, .
\ee
If we choose
\be
T(N) = \left ( \frac{1}{2[N+1]^F}  \right  )^{\half}\, ,
\ee
then the algebra reduces to $b_1 b_1^\dag + b_1^\dag b_1 =1$\, ,
which is the algebra of undeformed standard fermions.  Next we obtain  $[N,b_1]= [N, T(N) b] = T[N,b]= -b_1$ and similarly  $[N,b_1^\dag ]= b_1^\dag $. Consequently the transformed operators
$b_1, b_1^\dag$ satisfy 
\be
b_1 b_1^\dag + b_1^\dag b_1 =1, \quad [N,b_1]= -b_1, \quad [N,b_1^\dag] = b_1^\dag\, .
\ee
We can prove that this algebra is consistent with the exclusion principle: explicitly, 
\be
\{b_1, b_1 \} = \{ T b , Tb \}= T \{  b , T \}b - T T \{  b , b \} + \{ T  , T \}bb - T\{ T  , b \} b\, .
\ee
The first and the last terms cancel and the right hand side then 
vanishes identically.
This algebra is then no different from the one in Eq.(\ref{16}). Consequently we have shown that there is a transformation which can transform the deformation away, reducing the non-exclusion algebra to an algebra with exclusion, hence our conclusion is different from that of ref. \cite{Chai}.

We can prove another interesting result:  the deformed \q fermion algebra can be transformed into ordinary undeformed boson algebra, thus implying a \q fermion-boson transmutation. The desired transformation  is 
\be
b \rightarrow  \;  a_2= \left ( \frac{N+1}{[N+1]^F} \right  ) b, \quad b^\dag \rightarrow  \; a_2^\dag = b^\dag \left  ( \frac{N+1}{[N+1]^F}\right   )
\label{27}
\ee
is such a transformation. It can be  verified explicitly that  the operators $a_2, a_2^\dag$ satisfy the relations
\be
a_2 a^\dag_2 - a^\dag_2 a_2 =1, \quad [N,a_2]=-a_2, \quad [N, a_2^\dag] = a_2^\dag
\ee
and we also find $a_2 a_2^\dag = N+1 $ and $a_2 ^\dag a_2 = N$. This is thus identical with the algebra of ordinary bosons as in Eq.(\ref{1}).

The transformation which leads to the undeformed boson algebra is very intriguing. We shall conclude by pointing out  that the transformed operators $a_2, a_2^\dag$  satisfies all the expected properties. We know that the boson operators must obey $[a_2,a_2]=0, \; [a_2^\dag , a_2^\dag ]=0$. We can prove this explicitly as follows. From Eq.(\ref{27}) with $a_2=Tb= \left ( (N+1)/ [N+1]^F\right )^{\half}b$,  we immediately obtain
\br
[a_2,a_2]&=& [T b, \; Tb]\\
&=& T \{ b, \; T \} b - T T \{b, \; b \} + \{ T, \; T \} b b- T \{ T, \; b \} b\, .
\er
The first and fourth terms cancel out and we find
\be
[a_2,a_2]= - T^2 \{b, \; b\} + \{ T, \; T \} b b = -2 T^2 b b + 2 T^2 b b \equiv 0.
\ee
We do not require  $b^2=0$ and thus the above result is true for any $q$. Similarly $a_2^\dag$ commutes with itself. Finally the following results can be obtained immediately:
\be
[N,a_2] = [a_2^\dag a_2, a_2]= -a_2, \quad [N, a_2^{\dag}] = [a_2^\dag a_2, a_2^\dag ]= a_2^\dag
\ee
 Hence we have shown that all of the familiar standard properties of the boson algebra for $a_2, a_2^\dag$ follow while the operators $b, b^\dag$ satisfy the \q deformed fermion algebra, without requiring $b^2=0, (b^\dag )^2$.

In conclusion, we may state that the algebra of deformed fermions can be transformed to that of undeformed standard fermions as well as  to that of undeformed standard bosons. This indicates a \q fermi-\q bose transmutation. It would be worthwhile to investigate whether these transformations would allow an interpolation between Fermi and Bose distributions. It is interesting to note that Vokos {\it et al} have mentioned the possibility of \q fermions behaving like \q bosons \cite {Vokos} and Lutzenko {\it et al} have demonstrated  \cite {Lutz} the behavior of \q bosons as quasifermions.

\acknowledgments

 It is a pleasure  to thank Andrea Lavagno for   many useful discussions and for interest in this work.

\end{document}